\documentstyle{article}
\input epsf

\newcommand\be{\begin{equation}}
\newcommand\ee{\end{equation}}
\newcommand\bea{\begin{eqnarray}}
\newcommand\eea{\end{eqnarray}}
\newcommand\ket[1]{|#1\rangle}
\newcommand\bra[1]{\langle #1|}

\newcommand{\fatalpha}{{\bf \alpha \kern -0.44em \alpha}}
\newcommand{\fatsigma}{{\bf \sigma \kern -0.54em \sigma}}
\newcommand{\tpchi}{{\bf \chi \kern -0.35em \chi}}
\newcommand{\llambda}{{\bf \lambda \kern -0.45em \lambda}}
\begin{document}
\parbox{13 cm}
{
\begin{flushleft}
\vspace* {1.2 cm} {\large\bf {The quantum way to diagonalize
hermitean matrices}
}\\
\vskip 1truecm {\large\bf {Stefan Weigert}
}\\
\vskip 5truemm {HuMP -- Hull Mathematical Physics\\
                Department of Mathematics\\
                University of Hull\\
                Cottingham Road\\
                UK-Hull}

\end{flushleft}
}
\vskip 0.5truecm {\bf Abstract:}\\
An entirely quantum mechanical approach to diagonalize hermitean
matrices has been presented recently. Here, the genuinely quantum
mechanical approach is considered in detail for $(2\times2)$
matrices. The method is based on the measurement of quantum
mechanical observables which provides the computational resource.
In brief, quantum mechanics is able to directly address and output
eigenvalues of hermitean matrices. The simple low-dimensional case
allows one to illustrate the conceptual features of the general
method which applies to $(N \times N)$ hermitean matrices.
\hspace{4.3cm}({Fortschr. Phys. {\bf 51}, 248 (2003)}) \vskip 0.1
cm
\noindent

\subsection*{Introduction}
A new attitude towards quantum theory has emerged in recent years.
The focus is no longer on attempts to come to terms with
counter-intuitive quantum features but to capitalize on them. In
this way, surprising methods have been uncovered to solve specific
problems by means which have no classical equivalent: {\em quantum
cryptography}, for example, allows one to establish secure keys
for secret transmission of information \cite{bennet+84};
entanglement \cite{schroedinger35/2} is used as a tool to set up
powerful {\em quantum algorithms} which do factor large integers
much more efficiently than any presently known classical algorithm
\cite{shor95}. Throughout, these techniques make use of the {\em
measurement} of quantum mechanical observables as an unquestioned
tool. This is also true for many (but not all \cite{aharonov+99})
proposals of {\em quantum error correction} schemes
\cite{shor95,steane96}, required to let a potential algorithm run.

The purpose of the present contribution is to study the simplest
situation in which a quantum mechanical measurement, i.e. the bare
`projection' \cite{schroedinger35/2}, ``does'' the computation.
The computational task is to determine the eigenvalues of
hermitean $(2 \times 2)$ matrices by quantum means alone. Although
the answer to this problem can be given analytically, it is useful
to discuss this particular case since there is no conceptual
difference between diagonalizing $(2 \times 2)$ or $(N \times N)$
hermitean matrices along these lines \cite{weigert01}.

Before turning to the explicit example, consider briefly the
traditional view on quantum mechanical measurements: a measurement
is thought to confirm or reveal some information about the {\em
state} of the system. The measured observable ${\widehat A}$ is
assumed to be known entirely, including in-principle-knowledge of
its eigenstates and eigenvalues. Further, the observable also
defines the {\em scope} of the possible results of a measurement
since both the only allowed outcomes are its eigenvalues and,
directly after the measurement, the system necessarily resides in
the corresponding eigenstate.

In the context of quantum diagonalization, however, the crucial
idea is to learn something about the measured {\em
observable}---not about the state of the system. Why is there
scope for information gain at all? What can one learn from a
quantum mechanical experiment if both the measured observable and
the state are known?

It is essential to realize that the {\em input} required to
measure an observable $\widehat A$ and the {\em output} of an
experiment, in which $\widehat A$ is actually measured, are {\em
not} identical. In fact, it is possible to construct an apparatus
which measures an observable $\widehat A$ {\em without} explicitly
knowing its eigenvalues. Then, a measurement provides
partial---but explicit---information about the spectral properties
of the observable $\widehat A$: it delivers one of its
eigenvalues. As the eigenvalues of ${\widehat A}$ are {\em not}
known explicitly before the measurement, information is indeed
gained by measuring ${\widehat A}$. This is the idea which
underlies quantum diagonalization.

The quantum diagonalization of a hermitean matrix is achieved in
five steps: (1) express the matrix in a {\em standard form}; (2)
associate a quantum mechanical {\em observable} with it; (3)
identify an {\em apparatus} capable of measuring the observable;
(4) {\em measure} the observable---this provides the {\em
eigenvalues} of the matrix; (5) determine its {\em eigenstates}.
The next section gives the details for $(2 \times 2)$ matrices.
Then, the generalization to $(N \times N)$ matrices is briefly
summarized.

%
\subsection*{Quantum diagonalization of hermitean
             $(2 \times 2)$ matrices}
\label{twobytwo}
Suppose you want to determine the eigenvalues $A_\pm$ (and,
subsequently, the eigenvectors $\ket{A_\pm}$) of the general
hermitean $(2 \times 2)$ matrix
\be
{\sf A} =
\left[
\begin{array}{cc}
\alpha    &  \beta^* \\ \beta &  \gamma
\end{array}
\right] \, , \qquad \alpha, \gamma \in I \!\! R \, , \quad \beta
\in l\!\!\! C \, .
\label{22matrix} \ee
Obviously, this problem is easily solved analytically. The
eigenvalues read
\be \label{twoeigenvalues}
A_\pm = \frac{1}{2} \left( \alpha + \gamma
               \pm \sqrt{(\alpha - \gamma)^2 + 4 \beta \beta^*}
                     \right) \, ,
\ee
and one can also give explicit expressions for the eigenvectors of
the matrix ${\sf A}$. The five-step procedure of quantum
diagonalization is now applied to $\sf A$; it will, of course,
reproduce the result (\ref{twoeigenvalues}). However, all
conceptual points, which also apply to the technically more
cumbersome case of $(N \times N)$ matrices are conveniently
illustrated by this simple example.

\begin{enumerate}
  \item
\underline{Standard form of ${\sf A}$:} Any hermitean $(2\times
2)$ matrix ${\sf A}$ can be written as a unique linear combination
of the three Pauli matrices ${\sigma}_j = {\sigma}_j^\dagger, j=
1,2,3,$ , and the unit matrix ${\sigma}_0= {\sf I}_2$,
\be
{\sf A} =
\left( {\sf a}_0 {\sigma}_0 + {\sf a}  \cdot {\fatsigma} \right)
\, , \qquad {\sf a}_j = \frac{1}{2} \mbox{ Tr }\left[ {\sf A}
\sigma_j \right]             \in I  \! \! R \, ,
\label{defineM} \ee
where
\be \label{twocoeff}
 {\sf a}_0 = \frac{1}{2} ( \alpha + \gamma) \, , \qquad
 {\sf a}_1 = \frac{1}{2} ( \beta + \beta^*) \, , \qquad
 {\sf a}_2 = \frac{1}{2i} ( \beta^* - \beta) \, , \qquad
 {\sf a}_3 = \frac{1}{2} ( \alpha - \gamma) \, .
\ee
From a general point of view, this corresponds to an expansion of
${\sf A}$ in  multipole operators (cf. below).\\

\item \underline{Identification of an observable:} The most general
Hamiltonian of a spin-$1/2$ in a homogeneous magnetic field ${\bf
B}_0$ is linear in the components of the spin ${\sf S} =
\hbar{\mathbf \fatsigma}/2$. Therefore, any matrix ${\sf A}$ has
an interpretation as a Hamiltonian operator of a quantum spin
subjected to an appropriately chosen magnetic field,
\be
 {\sf A}   =      a {\sf I}_2 - g \mu_B {\bf B}_0 \cdot {\sf S}
           \equiv {\sf H}_{\sf A} ({\sf S}) \, , \qquad
         a = {\sf a}_0 \, , \quad
 {\bf B}_0 = \frac{- 2}{ g \mu_B \hbar}{\sf a} \, .
\label{hamilton} \ee
The part $a {\sf I}_2$ shifts the energy globally by a fixed
amount. Let $E_\pm$ denote the eigenvalues of the second part of
this expression, $ - g \mu_B {\bf B}_0 \cdot {\sf S} \equiv {\sf
H}^0_{\sf A} ({\sf S})$; then, the eigenvalues $A_\pm$ of the
matrix ${\sf A}$ are given by
\be
A_\pm = a + E_\pm \, .
\label{evrel} \ee
Consequently, the door is now open to determine the eigenvalues of
${\sf A}$ experimentally, i.e. through {\em measuring} the
eigenvalues $E_\pm$ of the Hamiltonian ${\sf H}^0_{\sf A} ({\sf
S})$. In the next step it is shown how to devise an apparatus
which measures this operator.\\

\item \underline{Setting up a measuring device:}
The apparatus is required to measure the eigenvalues of the
operator ${\sf H}^0_{\sf A} ({\sf S})$. In the case of a $( 2
\times 2)$ matrix this is just a familiar Stern-Gerlach apparatus,
appropriately oriented in space. However, as the method will be
applied to $(N \times N)$ hermitean matrices later on, it is
important to go through the construction of the measuring device
in detail.

Consider the spatially varying Hamiltonian
\be
{\sf H}^0 ({\bf r}, {\sf S}) = a {\sf I}_2 - g \mu_B {\bf B}({\bf
r}) \cdot {\sf S} \equiv a {\sf I}_2 + {\sf H}^0_{\sf A} ({\bf r},
{\sf S})\, ,
\label{spatialHamiltonian} \ee
which describes the interaction of a spin with an inhomogeneous
magnetic field ${\bf B}({\bf r})$. In order that ${\sf H}^0 ({\bf
r},{\sf S})$ measure the observable ${\sf H}_{\sf A}^0({\sf S})$,
the field needs to satisfy the conditions
\be
{\bf B}(0) = {\bf B}_0 \,  \quad \mbox{ and } \quad
 \nabla \cdot {\bf B}({\bf r}) = \nabla \times {\bf B}({\bf r})
 = 0 \, .
 \label{1/2field} \ee
Then, at the centre of the apparatus, the operator ${\sf H}^0_{\sf
A} ({\bf r}, {\sf S})$ coincides with the observable to be
measured, ${\sf H}^0_{\sf A} (0,{\sf S}) = {\sf H}_{\sf A}^0({\sf
S})$, and the magnetic field is consistent with Maxwell's
equations. Consider the field \cite{swift+80}
\be
{\bf B}({\bf r}) = (1+ {\bf k} \cdot {\bf r} ) {\bf B}_0 + ({\bf
B}_0 \cdot {\bf r}) {\bf k} \, ,
\label{Swiftchoice} \ee
which is consistent with Eqs. (\ref{1/2field}) if  the vector
${\bf k}$ is perpendicular to ${\bf B}_0$. Diagonalize the
operator ${\sf H}^0_{\sf A} ({\bf r},{\sf S})$ to first order of $
\nabla | {\bf B} |/ | {\bf B} |$. This leads to ${\bf
r}$-dependent eigenvalues
\be
E_\pm ({\bf r}) = \pm \frac{\hbar}{2} (1+ {\bf k} \cdot {\bf r} )
B_0\, ,
\label{rdepev} \ee
which imply the existence of a state-dependent force
\be
{\bf F}_\pm ({\bf r}) = - \nabla E_\pm ({\bf r})
  = \mp \frac{\hbar }{2} B_0 {\bf k} \ .
\label{1/2force} \ee
Consequently, particles on their way through the apparatus will be
deflected deterministically from a straight line once the
projection to an eigenstate $\ket{E_\pm}$ has occurred. In this
way, the eigenvalues of the observable can be accessed
experimentally.\\

\item \underline{Determination of the eigenvalues:}
If a measurement of the operator ${\sf H}^0_{\sf A}({\sf S})$ is
performed on the state $\hat \rho = {\sf I}_2 /2$, then it is
thrown with probability $1/2$ into one of the eigenstates
$\ket{E_\pm}$ with density matrix ${\hat \rho}_\pm$,
\be
\mbox{app}( {\sf H}^0_{\sf A}) : \quad
 \hat \rho = {\sf I}_2/2 \quad \stackrel{p_\pm}
 {\longrightarrow} \quad
 \left( E_\pm ; \, {\hat \rho}_\pm \, \right) \, , \qquad
 p_\pm = \mbox{ Tr } \left[\hat \rho {\hat \rho}_\pm
              \right] = \frac{1}{2}\, .
\label{measureH} \ee
Repeating the measurement a few times, both eigenvalues will have
been found soon. The probability not to obtain one of the two
values after $N_0$ identical runs of the experiment equals
$1/2^{N_0}$, and hence goes to zero exponentially with the number
of runs. Subsequently, the sought-after eigenvalues $A_\pm$ of the
matrix ${\sf A}$ are known due to the relation (\ref{evrel}), and
the major step in the diagonalization of the matrix has been
achieved in a quantum way.\\

\item \underline{Determination of the eigenstates:}
Once both eigenvalues $A_\pm$ are known, it is straightforward to
to determine the associated eigenstates by a simple calculation.
Optionally, one continues along an experimental line. One exploits
the fact that the apparatus app(${\sf H}_A^0$) prepares an
ensemble of eigenstates of ${\sf A}$ with density matrix ${\hat
\rho}_+$ if the other subbeam (containing ${\hat \rho}_-$) is
blocked---and {\em vice versa}. The Bloch representation of the
density matrix ${\hat \rho}_+$, say, can be parameterized by {\em
expectation values},
\be
{\hat \rho}_+
         = \frac{1}{2} \left( \sigma_0
           + \langle { {\fatsigma}} \rangle_+ \cdot {\fatsigma}\right) \, ,
\qquad \langle \hat \sigma_j \rangle_+ = \mbox{ Tr }\left[ {\hat
\rho}_+ {\sigma}_j \right] \equiv \bra{E_+} {\sigma}_j \ket{E_+}
\, .
\label{measurerho+} \ee
Hence, the components of the vector $\langle {\fatsigma}
\rangle_+$ (and therefore ${\hat \rho}_+$) are easily obtained by
means of a second, appropriately oriented Stern-Gerlach apparatus,
which amounts to a reconstruction of the density matrix ${\hat
\rho}_+$.
\end{enumerate}

\subsection*{Quantum diagonalization of hermitean $(N \times N)$
             matrices}
%
%
Five steps are necessary to diagonalize a hermitean  $(N \times
N)$ matrix ${\sf A}$ by quantum means.

\begin{enumerate}
  \item \underline{Standard form of ${\sf A}$:}
Write the hermitean $(N\times N$) matrix ${\sf A}$ as a
combination of linearly independent hermitean {\em multipole}
operators ${\sf T}_\nu, \nu= 0, \dots ,N^2-1,$
\be
{\sf A} = \sum_{\nu=0}^{N^2-1} {\bf a}_{\nu} {\sf T}_{\nu} \, ,
\qquad {\bf a}_{\nu} = \frac{1}{N}\mbox{ Tr } \left[{\sf A} \,
{\sf T}_{\nu} \right] \in I \!\! R \, .
\label{expandgen} \ee
Multipole operators ${\sf T}_{j_1j_2\cdots j_a}$ act in a Hilbert
space ${\cal H}_s$ of dimension $N= 2s+1$ which carries an
irreducible representation of the group $SU(2)$ with the spin
components $({\sf S_1},{\sf S_2},{\sf S_3})$ as generators. They
are defined as the symmetrized products ${\sf S}_{j_1} {\sf
S}_{j_2} \cdots {\sf S}_{j_a}, j_i = 1,2,3,$ and $a=0,1,\ldots,
2s,$ after subtracting off the trace, except for ${\sf T}_0 \equiv
{\sf T}^{(0)} = {\sf I}$, the $(N\times N)$ unit matrix. The index
$a$ labels $(2s+1)$ classes with $(2a+1)$ elements each,
transforming among themselves under rotations. Explicitly, the
lowest multipoles read
\be
 {\sf T}^{(1)}_{j} = {\sf S}_j
\, , \quad {\sf T}^{(2)}_{j_1 j_2} = \frac{1}{2} \left( {\sf
S}_{j_1} {\sf S}_{j_2} +{\sf S}_{j_2}{\sf S}_{j_1} \right) -
\frac{\delta_{i_1 j_2}}{3} {\sf S}_{j_1}{\sf S}_{j_2} \, .
\label{mpoles012} \ee
For the sake of brevity, a collective index $\nu \equiv
(a;j_1,\ldots,j_k)$ has been used in (\ref{expandgen}), taking on
the values $\nu = 0, 1, \ldots, N^2-1$.  The $N^2$ self-adjoint
multipole operators ${\sf T}_{\nu} = {\sf T}_{\nu}^\dagger$ form a
basis in the space of hermitean operators acting on the
$N$-dimensional Hilbert space ${\cal H}_s$ \cite{swift+80}. Two
multipoles are orthogonal with respect to a scalar product defined
as the trace of their product: $(1/N) \mbox{ Tr } \left[{\sf
T}_{\nu} {\sf T}_{\nu'} \right] = \delta_{\nu\nu'}.$\\

\item \underline{Identification of an observable:}
        Interpret the matrix ${\sf A}$ as an observable ${\sf H}_A$
        for a single quantum spin ${\sf S}$ with quantum number
        $s= (N-1)/2$,
\begin{equation}\label{observableha}
{\sf A} = \sum_{\nu=0}^{N^2-1} {\bf a}_{\nu} {\sf T}_{\nu} ({\sf
S}) \equiv  {\sf H}_A ({\sf S})  \, ,
\end{equation}
using the expression of the multipoles ${\sf T}_{\nu} ({\sf S})$
in terms of the components of a spin. Since the multipoles are
expressed explicitly as a function of the spin components not
exceeding the power $2s$, it is justified to consider them and,
{\em a fortiori}, the quantity ${\sf H}_A({\sf S})$ as an {\em
observable} for a spin $s$.\\

\item \underline{Setting up a measuring device:}
Swift and Wright \cite{swift+80} have shown  how to devise, in
principle, a {\em generalized} Stern-Gerlach apparatus which
measures any observable ${\sf H}_A({\sf S})$---just as a
traditional Stern-Gerlach apparatus measures the spin component
${\bf n} \cdot {{\sf S}}$ along the direction ${\bf n}$. The
construction requires that arbitrary static electric and magnetic
fields, consistent with Maxwell's equations, can be created in the
laboratory. To construct an apparatus app$({\sf H}_A)$ means to
identify a spin Hamiltonian ${\sf H} ({\bf r}, {\sf S})$ which
splits an incoming beam of particles with spin $s$ into subbeams
corresponding to the eigenvalues $A_n$. The most general
Hamiltonian acting on the Hilbert space ${\cal H}_s$ of a spin $s$
reads
\be
{\sf H} ({\bf r}, {\sf S})
        = \sum_{\nu=0}^{N^2-1} \Phi_\nu ({\bf r})
                           {\sf T}_{\nu} \, ,
\label{genhamiltonian} \ee
with traceless multipoles (except for $\nu=0$), and totally
symmetric expansion coefficients $\Phi_\nu ({\bf r}) \, (\equiv
\Phi^{(k)}_{j_1 j_2 \ldots j_k}({\bf r}))$. These functions, which
vary in space, generalize the magnetic field ${\bf B}({\bf r})$ in
(\ref{genhamiltonian}). Tune the corresponding electric and
magnetic fields in such a way that the coefficients $\Phi_\nu({\bf
r})$ and their first derivatives with respect to some spatial
direction, $r_1$, say, satisfy
\be
\left. \Phi_\nu ({\bf r})\right|_{{\bf r} = 0} = {\sf a}_\nu \, ,
\quad \mbox{ and } \quad \left. \frac{\partial \Phi_\nu ({\bf r}
)}{\partial r_1}\right|_{{\bf r} = 0} = {\sf a}_\nu \, .
\label{tune} \ee
As shown explicitly in \cite{swift+80}, this is always possible
with realistic fields satisfying Maxwell's equations. Then, the
Hamiltonian in (\ref{genhamiltonian}) has two important
properties:
\begin{enumerate}
 \item[(i)] At the origin, ${\bf r} = 0$, it coincides with the matrix
${\sf H} (0, {\sf S}) = {\sf H}_A ({\sf S})$.
 \item[(ii)] Suppose
that a beam of particles with spin $s$ enters the generalized
Stern-Gerlach apparatus app(${\sf H}_A$) just described. At its
center, particles in an eigenstate $\ket{A_n}$, say, will
experience a force in the $r_1$ direction given (up to second
order in distance from the center) by
\be
\left. F_1 ({\bf  r}) \right|_{{\bf r} = 0} = - \left.
\frac{\partial  \bra{A_n}{\sf H} ({\bf r} , {\sf S}) \ket{A_n}}
{\partial r_1} \right|_{{\bf r} = 0} = - A_n \, , \qquad n= 1,
\ldots, 2s+1\, .
\label{separation} \ee
\end{enumerate}
Consequently, particles with a spin residing in different
eigenstates $\ket{A_n}$ of the operator ${\sf H}_A$ will be
separated spatially by this apparatus. From a conceptual point of
view, the procedure is equivalent to the method outlined above for
a spin $1/2$.\\

\item\underline{Determination of the eigenvalues:}
Once the apparatus app$({\sf H}_A)$ has been set up, one needs to
carry out measurements on (particles carrying) a spin $s$ prepared
in the homogeneous mixture ${\hat \rho} = {\sf I}_N /N$. The
output of each individual measurement will be one of the
eigenvalues $A_n$ of the matrix ${\sf A}$, according to the
`projection postulate:'
\be
\mbox{app}{({\hat A})}: \quad \hat \rho = {\sf I}_N /N \quad
\stackrel{p_n} {\longrightarrow} \quad \left( A_n ; {\hat \rho}_n
\right) \, , \qquad p_n = \mbox{ Tr } \left[ \hat \rho {\hat
\rho}_n \right] = \frac{1}{N} \, .
\label{measure} \ee
In words, the action of the apparatus is, with probability
$p_n=1/N$, to throw the system prepared in state $\hat \rho$ into
an eigenstate ${\hat \rho}_n \equiv \ket{A_n} \bra{A_n}$ of the
observable $\hat A$; the {\em outcome} of the measurement is given
by the associated eigenvalue $A_n$.

After sufficiently many repetitions, all eigenvalues will be
known, although the outcome of an {\em individual} measurement
cannot be predicted due to the probabilistic character of quantum
mechanics. It is necessary to repeat the experiment a number of
times until {\em all} values $A_n$ have been obtained. Since the
spin $s$ has been prepared initially in a homogeneous mixture,
$\hat {\rho} = {\sf I}_N /N$, the $(2s+1)$ possible outcomes occur
with equal probability. The probability {\em not} to have obtained
one specific value $A_n $ after $N_0 \gg N$ measurements equals
$(2s/(2s+1))^{N_0} \simeq \exp[-N_0/2s]$, decreasing exponentially
with $N_0$.

\item \underline{Determination of the eigenstates:} As before it is
possible to either calculate the eigenstates $\ket{A_n}$ of the
matrix ${\sf A}$ on the basis of the known eigenvalues, or to
determine them {\em experimentally} by methods of state
reconstruction (see \cite{weigert99/1} for details).
\end{enumerate}

\subsection*{Summary and Outlook}

As a result of the five steps just described, a hermitean $(N
\times N)$ matrix ${\sf A}$ has been diagonalized {\em without}
calculating the zeroes of its characteristic polynomial by
traditional means. The fourth step solves the hard part of the
eigenvalue problem since it provides the eigenvalues $A_n$ of the
matrix ${\sf A}$---information which cannot be obtained in closed
form if $N \geq 5$. One might best describe the measuring device
app(${\sf H}_A$) as a `special purpose machine' which is based on
the `collapse of the wave function' as computational resource.
However, one could avoid to use the notion of `collapse' or
`projection' by characterizing the process indirectly using the
concept of `repeatable measurements' described in \cite{peres95}.

By construction, the quantum mechanical diagonalization is not
based on the representation of a mathematical equation in terms of
a physical system which then would `simulate' it. Similarly, no
`software program' runs which would implement an diagonalization
algorithm. Therefore, the method resembles neither an analog nor a
digital calculation.

t is worthwhile to point out that the quantum mechanical approach
to the diagonalization of hermitean matrices is based on the
assumption that the behaviour of a spin $s$ is described correctly
by non-relativistic quantum mechanics. Note, finally, that quantum
diagonalization does {\em not} depend on a particular
interpretation of quantum mechanics.


\end{document}